\documentstyle[psfig,epsf,conf-X,10pt]{article}
\begin{document} 
\small
\heading{%
%Begin Heading
%
Evolution of the cluster X-ray luminosity function from 
the WARPS survey
}
\par\medskip\noindent
\author{%
%Begin Author names
L.R. Jones$^{1}$, H. Ebeling$^{2}$,  C. Scharf$^{3}$, E. Perlman$^{4}$,
D. Horner$^{5}$, B. Fairley$^{1}$, G. Wegner$^{6}$, M. Malkan$^{7}$
}
\address{%
%First address
School of Physics \& Astronomy, University of Birmingham,
Birmingham B15 2TT, UK.
}
\address{%
Institute for Astronomy, 2680 Woodlawn Dr, Honolulu, HI
96822, USA}
\address{%
Space Telescope Science Institute, Baltimore, MD 21218,
USA.}
\address{%
Dept of Physics and Astronomy, 
Johns Hopkins University,
%3400 North Charles Street,               
Baltimore, MD  21218, USA.}
\address{%
Lab for High Energy Astrophysics, Code 660,
NASA/GSFC, Greenbelt, MD 20771, USA.}
\address{%
Dept. of Physics \& Astronomy, Dartmouth College, 
% 6127 Wilder Lab., 
Hanover, NH 03755, USA.}
\address{%
Dept. of Astronomy, UCLA, Los Angeles, CA 90024, USA.}

\begin{abstract}
The  evolution of the X-ray luminosity function of clusters of galaxies
has been measured to z=0.85 using over 150 X-ray selected 
clusters found in the WARPS survey.
%The preliminary results are that 
We find no evidence for evolution 
of the luminosity function at any
luminosity or redshift. 
%Because of the large number of clusters in the survey, 
The observations constrain the evolution of the 
space density of moderate luminosity
clusters to be very small, and much less than predicted by most 
models of the growth of structure with $\Omega_m$=1.

%We also briefly summarize the results 

\end{abstract}
\section{Introduction}
The evolution of the space density of clusters of galaxies is a 
measurement sensitive
to the physical and cosmological parameters of models of structure 
formation. We describe a deep X-ray survey of clusters of galaxies 
(the Wide Angle ROSAT Pointed Survey or WARPS),
and use it to measure the evolution of the 
%X-ray survey designed to measure the evolution of the 
X-ray luminosity function (XLF) of clusters of galaxies
\cite{S97}, \cite{J98}, \cite{E99}).
Fairley et al (these proceedings) describe our results on the 
evolution of the cluster 
X-ray luminosity-temperature relation at high redshifts \cite{F00}.

\section{Survey method}
Our goal was to compile a complete, unbiased, X-ray selected sample
of clusters of galaxies from serendipitous detections in deep,
high-latitude ROSAT
PSPC pointings. A detailed review of the X-ray source detection 
algorithm used (Voronoi Tessellation and Percolation or VTP), 
the sample selection and flux correction techniques are given in 
\cite{S97}. VTP is particularly well suited to the 
detection and characterization of low surface brightness emission
and to the recognition of extended sources. Only the centre of
the ROSAT PSPC field was used, up to an
off-axis angle of 15 arcmin. However, even at this relatively small
off-axis angle, the PSPC point-spread function degrades to 
$\approx$55 arcsec (FWHM, at 1 keV) and there is a possibility that 
some clusters at the edge of the PSPC fields (where most of the 
survey area is) would remain unresolved.
To reduce this possible incompleteness, our optical follow-up 
observations are not limited to extended X-ray sources but also 
include likely point X-ray sources without obvious optical counterparts.
\noindent
The flux limit is 5.5x10$^{-14}$ erg cm$^{-2}$ s$^{-1}$ (0.5-2 keV)
and the total solid angle 73 deg$^2$. 
Detailed simulations were performed
to derive the 
survey sensitivity as a function of both source flux and source extent.

\begin{figure}
\mbox{\epsfxsize=7.2cm \epsffile{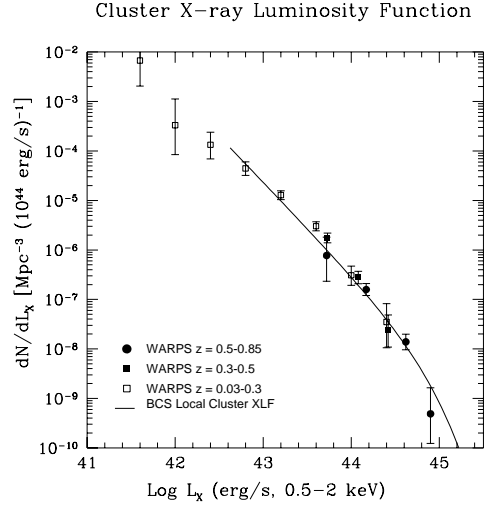}}
\mbox{\epsfxsize=7.2cm \epsffile{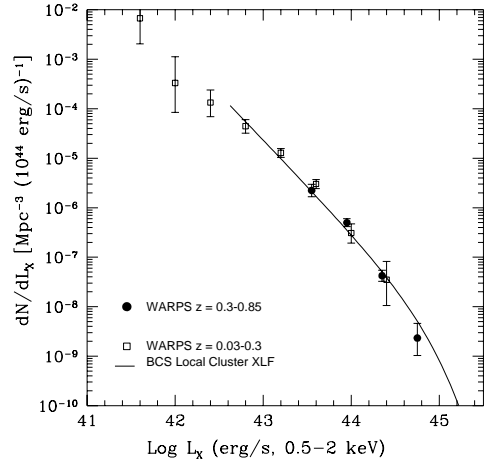}}
%
%\begin{figure}
%\centerline{\vbox{
%\psfig{figure=fig1.ps,height=7.cm}
%}}
\caption[]{
X-ray luminosity functions of clusters of galaxies. No evolution of
the space density of clusters is observed up to redshifts of z=0.85. 
The lower figure contains the same data as the upper figure, but 
combined into one redshift bin for z$>$0.3, in order to emphasize the
small statistical errors.
}
\end{figure}

\section{The X-ray luminosity function of clusters}
The survey currently contains over 150 clusters and groups of galaxies,
approximately half of which are at redshifts z$>$0.3. Spectroscopic
redshifts for all but a few have been obtained and the survey
is complete at z$<$0.85. The remaining few distant cluster candidates
are very probably all at higher redshifts, based on their brightest
cluster galaxy magnitudes. 
The XLFs shown here are based on an initial analysis; while some details
will change (eg correction for AGN contamination) in a final version,
we do not expect there to be major changes.

\noindent
The cluster XLF is shown in Fig 1. 
A value of q$_0$=0.5 has been assumed in calculating the luminosities
and volumes but the results are 
qualitatively similar for q$_0$=0.1.
The WARPS data points extend over almost 3 decades of luminosity,
from 10$^{42}$ erg s$^{-1}$
at low redshifts and up to 8x10$^{44}$ erg s$^{-1}$ (0.5-2 keV)
at high redshifts. There is no evidence for evolution of the 
XLF at any redshift up to z=0.8 when compared with the local BCS XLF 
\cite{E97}, shown as a solid line.
The error bars are based on a Poissonian distribution of the 
number of clusters in each bin. The survey area for each cluster has been 
calculated using the cluster flux but making the simplifying assumption
that all clusters have the same observed surface brightness profile 
(characterised as a constant angular core radius and $\beta$=2/3 index).
Mean differences from the true survey areas are small ($<$10\%).

As the XLFs at z=0.3-0.5 and z=0.5-0.85 are consistent with each other, in 
the lower panel of the figure we
have combined both redshift ranges to increase the statistical
accuracy at the expense of a rather broad redshift bin.  Nevertheless,
the small error bars on the high redshift points (filled circles) 
indicate that any evolutionary factor $>$1.5 in the space density of moderate 
luminosity clusters (3x10$^{43}<$L$_X<$3x10$^{44}$ erg s$^{-1}$)  is ruled out.

\section{Discussion}

Several recent deep X-ray clusters surveys are described in these proceedings 
and elsewhere. In the regime where these surveys have good statistical
accuracy (ie. moderate X-ray luminosities $\sim$3x10$^{43}$-3x10$^{44}$ 
erg s$^{-1}$) there is a consensus that {\it no evolution of the 
XLF is observed to z$\approx$0.7}. Five surveys agree on this point 
(EMSS, RDCS, SHARC, CfA, WARPS) and the only disagreement  
(the RIXOS survey \cite{C95}) can be understood in terms of 
the RIXOS source detection algorithm.

\noindent
Most models of the growth of structure in the Universe with $\Omega_m$=1 
predict rapid evolution of the space density of clusters of galaxies
(eg. scaling relation models \cite{B97} or a cold+hot dark
matter model \cite{B94}). The tight
constraints on the level of evolution allowed by the observations 
rules out most of the $\Omega_m$=1 models. In models with $\Omega_m\approx$0.3
(with or without a cosmological constant) the predicted evolution is much 
less rapid and these values of  $\Omega_m$ are consistent with the 
lack of evolution observed.

\noindent
At the higher X-ray luminosities of the most massive clusters 
(L$_X>$5x10$^{44}$ erg s$^{-1}$), there
is some disagreement between the results of different surveys as to
the degree of evolution found (if any), but this
may partly be due to the small numbers of high luminosity clusters in any 
one survey. This is unfortunate, since it is the most massive 
clusters which have most leverage to constrain the cosmological
parameters of model predictions. 
Wide area X-ray surveys, designed to detect the rare high luminosity,
massive clusters in large numbers at high redshifts (z$>$0.5) are planned 
(see Lumb \& Jones and Ebeling et al in these proceedings).

\noindent
Future work will concentrate on the handful of clusters in the 
WARPS survey at z$\approx$1, on measurements of the evolution of the   
cluster X-ray temperature function using Chandra and XMM, and studies 
of the Butcher-Oemler effect in X-ray selected clusters.  
%We are also studying the WARPS sample of `fossil' groups of galaxies,
%to help understand the formation of luminous elliptical galaxies
%\cite{J00}.

\acknowledgements{We acknowledge useful discussions with Pat Henry.
}

\vfill
\end{document}